\documentclass[a4paper]{spie}  
\usepackage[utf8x]{inputenc} 
\usepackage[]{graphicx}
\usepackage{subfigure}
\usepackage{transparent}
\usepackage{amssymb}
\usepackage{amsmath} 
\usepackage[english]{babel}
\usepackage{multicol}

\newcommand{\e}[1]{\times 10^{#1}} 
\newcommand{\ml}[1]{\mathrm{#1}} 
\newcommand{\ex}[1]{\mathrm{e}^{#1}} 
\newcommand{\dg}{^{\circ}}
\newcommand{\specialcell}[2][l]{%
  \begin{tabular}[#1]{@{}l@{}}#2\end{tabular}}

\title{First experimental results of very high accuracy centroiding measurements for the neat astrometric mission} 

\author{A. Crouzier\supit{a}, F. Malbet\supit{a}, O. Preis\supit{a}, F. Henault\supit{a}, P. Kern\supit{a}, G. Martin\supit{a}, P. Feautrier\supit{a}, E. Stadler\supit{a}, S. Lafrasse\supit{a}, A. Delboulbe\supit{a}, E. Behar\supit{a}, M. Saint-Pe\supit{a}, J. Dupont\supit{a}, S. Potin\supit{a}, C. Cara\supit{b}, M. Donati\supit{b}, E. Doumayrou\supit{b}, P. O. Lagage\supit{b}, A. Léger\supit{c}, J. M. LeDuigou\supit{d}, M. Shao\supit{e}, R. Goullioud\supit{e}
\skiplinehalf
\supit{a}Institut d'Astrophysique et de Planétologie de Grenoble, 414 Rue de la Piscine, St Martin d'Hères, Grenoble, France; \\
\supit{b}Commisariat à l'Energie Atomique et aux Energies Alternatives, Saclay, centre d'études nucléaires de Saclay, Paris, France; \\
\supit{c}Institut d'Astrophysique Spatiale, Centre universitaire d'Orsay, Paris, France; \\
\supit{d}Centre National d'Etudes Statiales, 2 place Maurice Quentin, Paris, France; \\
\supit{e}Jet Propulsion Laboratory, 4800 Oak Grove Drive, Pasadena, CA, U.S.A. 91109
}

\authorinfo{Further author information: (Send correspondence to A.A.A.)\\A.A.A.: E-mail: aaa@tbk2.edu, Telephone: 1 505 123 1234\\  B.B.A.: E-mail: bba@cmp.com, Telephone: +33 (0)1 98 76 54 32}

\begin{document} 
\maketitle 
\graphicspath{ {./Figures/} }
\begin{abstract}

NEAT is an astrometric mission proposed to ESA with the objectives of detecting Earth-like exoplanets in the habitable zone of nearby solar-type stars. NEAT requires the capability to measure stellar centroids at the precision of $5\e{-6}$ pixel. Current state-of-the-art methods for centroid estimation have reached a precision of about $2\e{-5}$ pixel at two times Nyquist sampling, this was shown at the JPL by the VESTA experiment. A metrology system was used to calibrate intra and inter pixel quantum efficiency variations in order to correct pixelation errors.

The European part of the NEAT consortium is building a testbed in vacuum in order to achieve $5\e{-6}$ pixel precision for the centroid estimation. The goal is to provide a proof of concept for the precision requirement of the NEAT spacecraft. In this paper we present the metrology and the pseudo stellar sources sub-systems, we present a performance model and an error budget of the experiment and we report the present status of the demonstration. Finally we also present our first results: the experiment had its first light in July 2013 and a first set of data was taken in air. The analysis of this first set of data showed that we can already measure the pixel positions with an accuracy of about $1\e{-4}$ pixel.


\end{abstract}


\keywords{exoplanets, astrometry, space telescope, centroid, calibration, micro-pixel accuracy, interferometry, metrology, data processing}


\section{INTRODUCTION}\label{sec:INTRODUCTION} 

\subsection{Presentation of the NEAT concept}\label{subsec:Presentation of the NEAT concept}

With the present state of exoplanet detection techniques, none of the rocky planets of the Solar System would be detected and indeed their presence is a very strong constraint on the scenarios of the formation of planetary systems. By measuring the reflex motion of planets on their central host stars, astrometry can yield to the mass of planets and to their orbit determination. This technique is used frequently in binary studies and is very successful to determine the masses and the orbits of multiple stars. However it is necessary to go to space to reach the precision required to detect all planets down to the telluric regime.
\\

\begin{figure}[t]
\begin{center}
\includegraphics[height = 40mm]{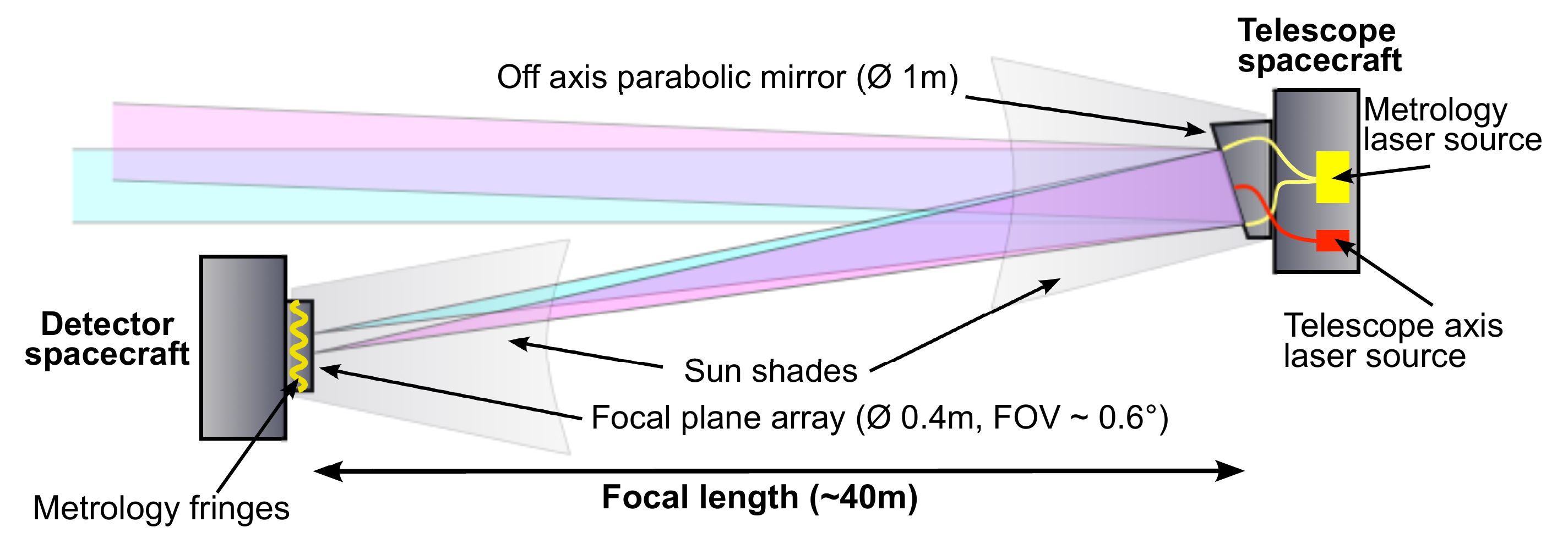}
\caption{\label{neat_concept_diagram}\textbf{The proposed NEAT concept.} The metrology system projects dynamic Young fringes on the detector plane. The fringes allow a very precise calibration of the CCD in order to reach micro-pixel centroiding errors.}
\end{center}
\end{figure}

We are proposing a mission to ESA in the framework of the call for M missions in the Cosmic Vision plan which objective is to find most of the exoplanets of our Solar neighbourhood\cite{Malbet11,Malbet12,Malbet13}. The objective is to use differential astrometry to complete the measurements obtained by other techniques in order to lower the threshold of detection and characterization down to the level of an Earth mass in the habitable zone of each system. We want to explore in a systematic manner all solar-type stars (FGK spectral type) up to 20 pc from the Sun. The satellite concept is based on formation flying technology with a satellite carrying a single primary mirror and another satellite carrying the focal plane (see Fig.~\ref{neat_concept_diagram}). The measurement is done using laser metrology and interferometry.

One of the fundamental aspects of the NEAT mission is the extremely high precision required to detect exo-Earths in habitable zone by astrometry. The amplitude of the astrometric signal that a planet leaves on its host star is given by the following formula:
\begin{equation}\label{eq:astrometric_signal}
A = 3 \mu \ml{as} \times \frac{M_{\ml{Planet}}}{M_{\ml{Earth}}} \times \left(\frac{M_{\ml{Star}}}{M_{\ml{Sun}}}\right)^{-1} \times \frac{R}{1\ml{AU}} \times \left(\frac{D}{1\ml{pc}}\right)^{-1}
\end{equation}

Where $D$ is the distance between the sun and the observed star, $M_{\ml{Planet}}$ is the exoplanet mass, $R$ is the exoplanet semi major axis and $M_{\ml{Star}}$ is the mass of the observed host star. For an Earth in the habitable zone located at 10 pc from the sun, the astrometric signal is 0.3 micro arcseconds (or $1.45\e{-11}$ rad). This is smaller than the precision announced for the Gaia mission (launch scheduled for 2013) which should be 7 $\mu$as, in optimal conditions. With a focal length of 40 meters, and taking into account a required signal to noise ratio\cite{sim_double_blind_test09} of 6 and the required number of measurements per target \cite{neat_number_of_measurements11}, the 0.3 $\mu$as requirement to detect an Earth at 10 pc translates into a need to calibrate the pixelation error to $5\e{-6}$ pixels for each integration, as shown by the NEAT error budget\cite{neat_error_budget11}.
\\

For more details about the NEAT mission we invite the reader to refer to the other paper of the same volume by Fabien Malbet\cite{Malbet13}. In the following subsections we will present the CNES centroid experiment, which goal is to demonstrate the feasibility of the NEAT requirement. In the last section we report our progress and the first results of the experiment.

\subsubsection{CNES centroid experiment: context and timeline}

In order to strengthen the NEAT case in the next ESA call for M class missions, the European part of the NEAT consortium is designing and building a testbed very similar to the one that is used at the JPL\cite{Nemati11}. The main difference is that the metrology system will be made of integrated photonic components. The design of the testbed has begun on January 2012 and the components have now been procured. We have obtained first light in 2013 (see section \ref{sec:CURRENT STATUS OF THE TESTBED AND LAST RESULTS}) and we expect to reach final performance in 2014.
\\

The laboratories involved in this project are: IPAG (Institut de Planétologie et d'Astrophysique de Grenoble) - the laboratory where the experiment takes place, CEA (Commisariat à l'Energie Atomique et aux Energies Alternatives) where the electronics for the CCD camera has been developed, IAS (Institut d'Astrophysique Spatiale), JPL (Jet Propulsion Laboratory) from which the past experience is very valuable to us.
\\

The founding is brought by the CNES (Centre National d'Etudes spatiales) and the labex OSUG@2020.

\subsubsection{CNES centroid experiment: presentation of the testbed}

\begin{figure}[t]
\begin{center}
\subfigure[]{\label{fig:centroid_xp_diagram}
\includegraphics[width=7cm]{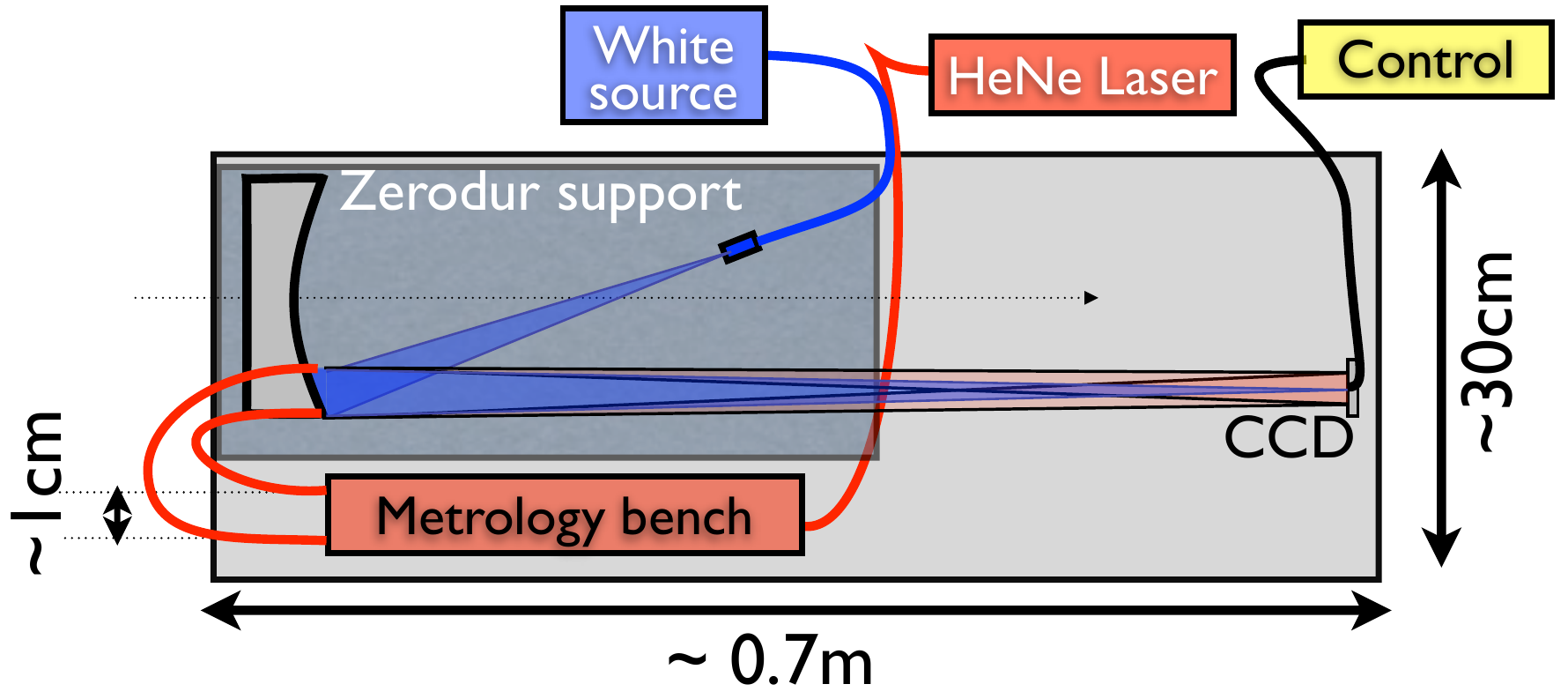}}
\hspace{5pt}
\subfigure[]{\label{fig:centroid_xp_optical_setup}
\includegraphics[width=9cm]{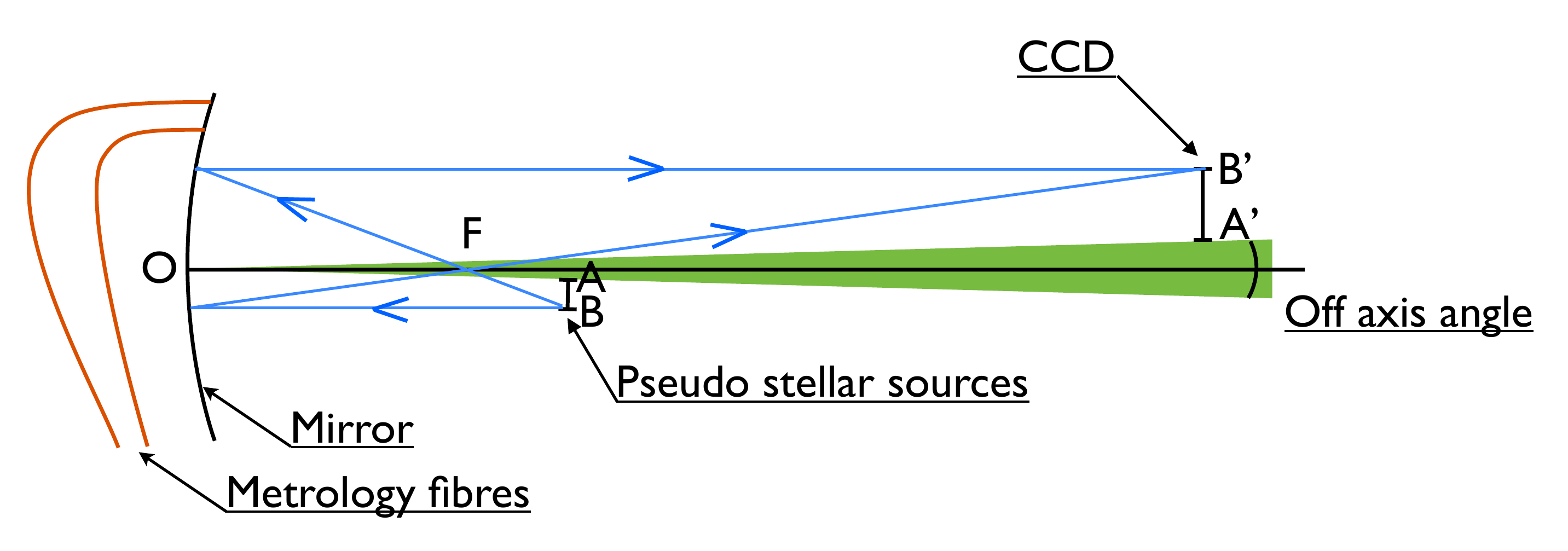}}
\end{center}
\caption{\label{JPL_xp}\textbf{Left: Schematic of the system's components. Right: The optical set-up of the experiment.}}
\end{figure}

\begin{table}[t]
\hspace{1cm}
\caption{\label{tab:notations}Notations.}
\begin{center}
\begin{tabular}{|l|l|l|}
  \hline
  Parameter & Notation & Value\\
  \hline
distance mirror to CCD (OA') & $L$ & 0.6 m\\
minimum/maximum wavelength of the pseudo stellar sources & $\lambda_{\ml{min}} / \lambda_{\ml{max}}$ & 0.4/0.8 µm\\
diameter of the entrance pupil & $D$ & 5 mm\\
mirror focal length (OF) & $f$ & 200 mm\\
separation between the pseudo stellar sources (AB) & $s$ & 240 µm\\
wavelength used for the metrology & $\lambda_m$ & 633 nm\\
metrology baseline & $B$ & 5 to 8 mm\\
  \hline
\end{tabular}
\end{center}
\end{table}

The testbed is a simple optical bench that mimics the NEAT optical layout. A spheric mirror images five pinholes which are illuminated by a white source onto a CCD, so that the image is diffraction limited. The five pinholes represent stars, we will refer to them as ‘‘pseudo stellar sources". A set of single-mode fibres located at the edge of the mirror produce laser fringes on the detector. A schematic of the system's components is shown in Fig.~\ref{fig:centroid_xp_diagram}. The optical set-up inside the vacuum chamber is shown in the Fig.~\ref{fig:centroid_xp_optical_setup}.
\\

The most innovative aspect of this experiment is the metrology system that will allow the micro-pixel calibration of the CCD. This system consist of at least two metrology bases (i.e the two pairs of single mode fibres), respectively aligned along the horizontal and vertical axis. The fibre extremities are located next to the mirror and project Young fringes on the detector. Additionally a phase modulator is used to dynamically sweep the fringes over the focal plane. By measuring the intensities variations of the signal for each pixel, one can characterise the inter and intra pixel response of the CCD and bring the centroid error down to the level of a few micro-pixels\cite{Zhai11}.
\\

In the next sections we present the design of the NEAT testbed, a performance model and the latest results. For all these sections we will use the notations of the Table~\ref{tab:notations}. The specifications of the testbed were presented in the proceedings of 2012\cite{Crouzier12}.

\section{DESIGN OF THE TESTBED}

\subsection{Bench and mechanical systems}\label{subsec:Bench and mechanical systems}

\begin{figure}[t]
\begin{center}
\includegraphics[height = 60mm]{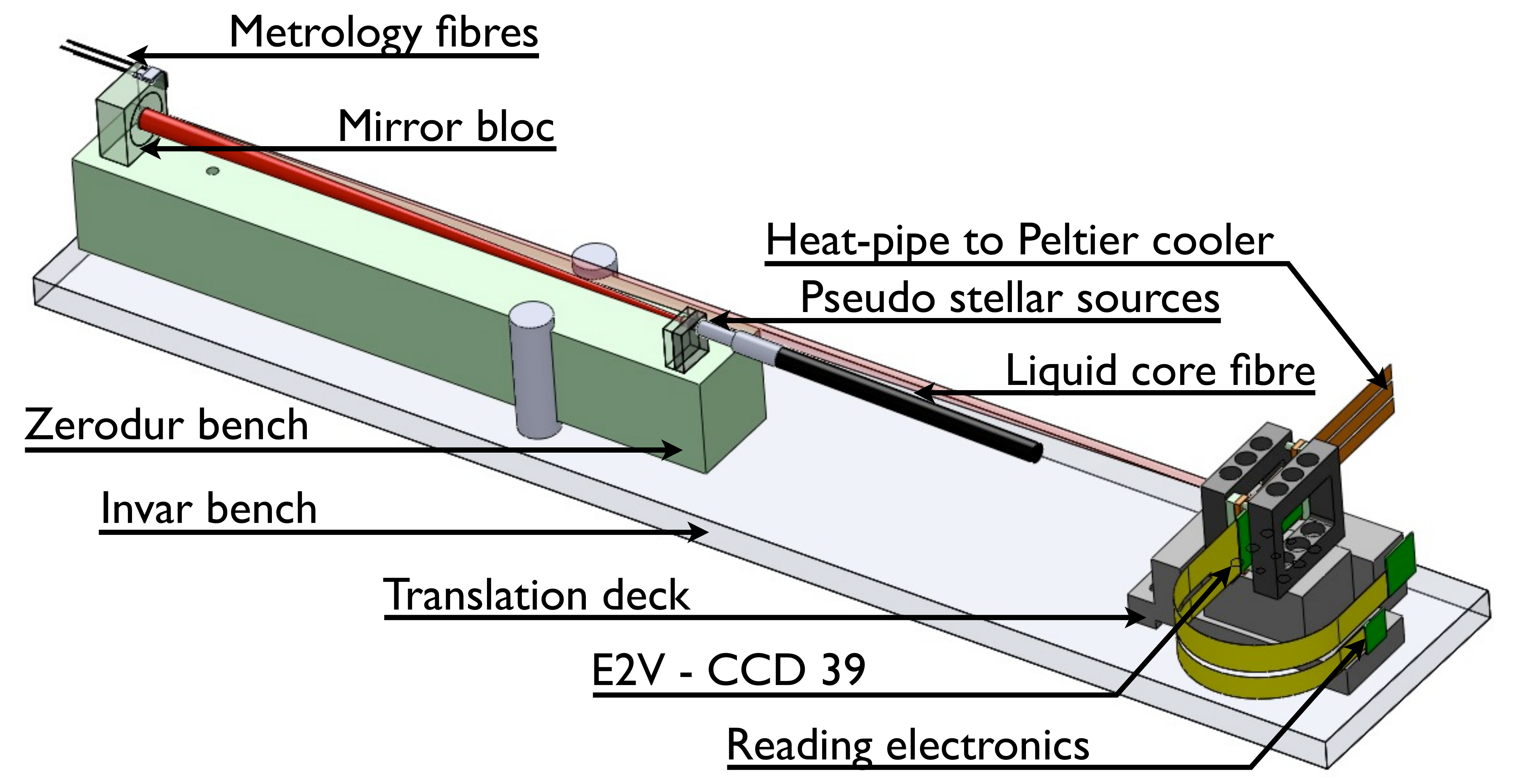}
\caption{\label{fig:centroid_xp_mechanical_concept}\textbf{3D view of the testbed.}}
\end{center}
\end{figure}

Figure~\ref{fig:centroid_xp_mechanical_concept} shows a solidworks 3D view of the testbed. The metrology is presented in section \ref{subsec:Metrology}, the pseudo stellar sources are presented in section \ref{subsec:Pseudo stellar sources}.

The current design approach is based on prior experience at the JPL and on a best effort approach, within the allowed budget. The most critical part, i.e. the one that supports the CCD and the pseudo stellar sources (see Fig.\ref{fig:centroid_xp_mechanical_concept}) is made entirely out of zerodur. Additionally the bench will be thermally regulated to about 0.1 degree. The non critical part supporting the CCD can accommodate larger thermal expansion and will be made in Invar. The vacuum chamber containing the experiment is placed on a standard suspension table with passive pneumatic suspensions.

\subsection{Metrology}\label{subsec:Metrology}

\begin{figure}[t]
\begin{center}
\includegraphics[width = 15cm]{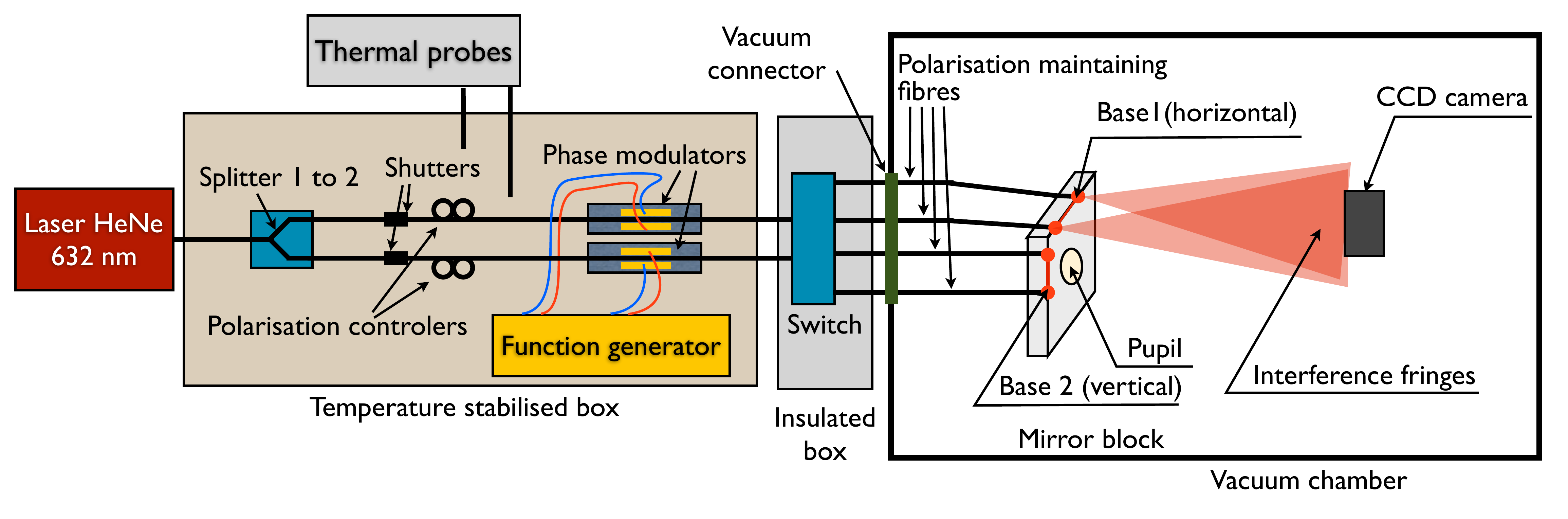}
\caption{\label{fig:metrology_v2}\textbf{Schematic of the metrology.}
}
\end{center}
\end{figure}

The metrology, which is made of integrated components, from the laser to the bases, is shown by Fig.~\ref{fig:metrology_v2}. The source for the metrology is a stabilised HeNe laser with a power output of 1.5mW. The frequency stability of the laser is 2 MHz (relative stability of $\delta \lambda / \lambda = 4\e{-9}$). The light from the laser is fed into the lithium niobate modulators to apply a periodic phase shift between the two lanes. This configuration ensures that the phase modulation is applied between the two sources constituting each base. The shutters are controlled to alternatively block all the metrology sources, during the pseudo stellar source integration phase, or to project either vertical or horizontal dynamic Young fringes during the CCD calibration phase.

\begin{figure}[t]
\begin{center}
\includegraphics[height = 8cm]{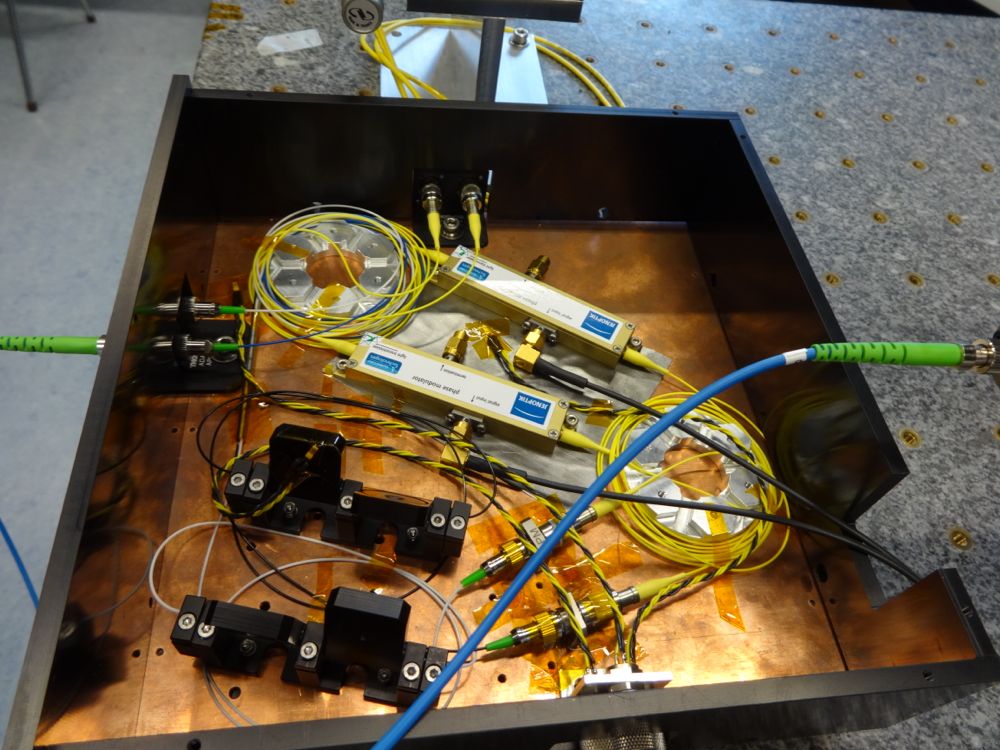}
\caption{\label{fig:metrology_box}\textbf{Picture of the metrology box.} This is the box between the laser and the switch. We can see the following components in the box: the numerous connectors, the two modulators and the polarisation controllers (Lefevre-loops).
}
\end{center}
\end{figure}

The interference pattern created by the horizontal metrology baseline:
\begin{equation}\label{interference_pattern}
I(x,y) = 2I_0\left[1+V\cos\left(\phi_0 + \Delta \phi (t) + \frac{2\pi xB}{\lambda_{\ml{met}} L}\right)\right]
\end{equation}
Where $I_0$ is the average intensity at the focal plane, $B$ is the metrology baseline, $L$ is the distance between the fibres and the CCD, $\phi_0$ is a static phase difference, $\Delta \phi (t)$ is the modulation applied between the lines and x is the horizontal spatial coordinate (i.e the one aligned with the metrology baseline). Although the exact shape of the fringes is hyperbolic, at the first order the fringes are straight and aligned with the direction perpendicular to the metrology baseline.
\\

If we assume that the sources are of equal intensity and that the intensity created at the focal plane is uniform we have V=1. But in reality, the visibility of the fringes is affected by the intensity mismatch and the polarisation between the sources. If we let $I_a$ and $I_b$ be the intensity of each source, and $\theta$ the angle between the polarisations (we assume completely linearly polarised sources), the visibility observed on the CCD is:
\begin{equation}\label{fringes_visibility}
V_{I(x,y)} = V_{\ml{pixel}} \times V_{\theta} \times V_{I(x,y)} = V_{\ml{pixel}} \times \frac{2 \sqrt{I_a(x,y) I_b(x,y)}}{I_a(x,y) + I_b(x,y)} \times \cos(\theta) 
\end{equation}

$V_{\ml{pixel}}$ is another visibility term that depends on the ratio pixel/fringe width. Because the width of the pixels is not negligible compared to the fringe size, the visibility on the CCD is lowered. This term is pixel specific as each pixel can have a different width.

To ensure a low noise on the relative phase between each metrology lane, it is important to avoid differential temperature variations between the lanes. That's why after the splitter, all the elements are enclosed in temperature stabilised boxes. The metrology box is shown on figure \ref{fig:metrology_box}.

\subsection{Pseudo stellar sources}\label{subsec:Pseudo stellar sources}

\begin{figure}[t]
\begin{center}
\includegraphics[height = 50mm]{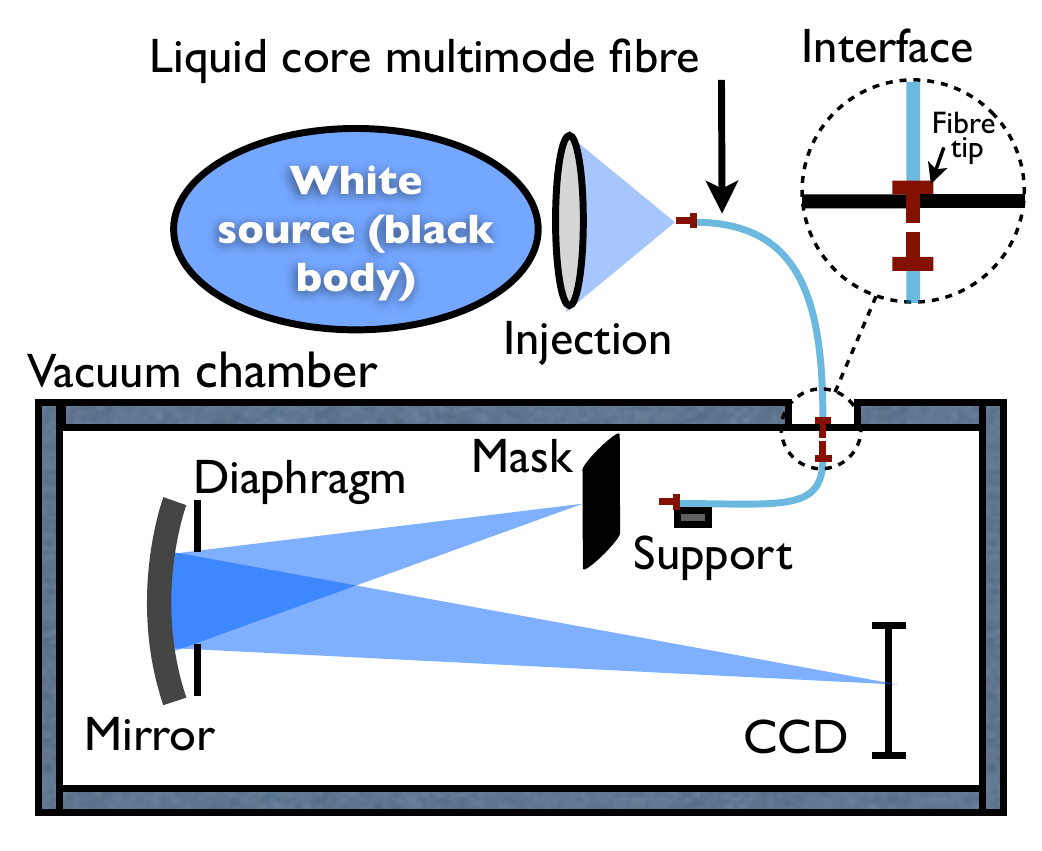}
\caption{\label{fig:scheme_pseudostellar_sources}\textbf{Schematic of the pseudo stellar sources.}}
\end{center}
\end{figure}

\begin{figure}[t]
\begin{center}
\includegraphics[height = 60mm]{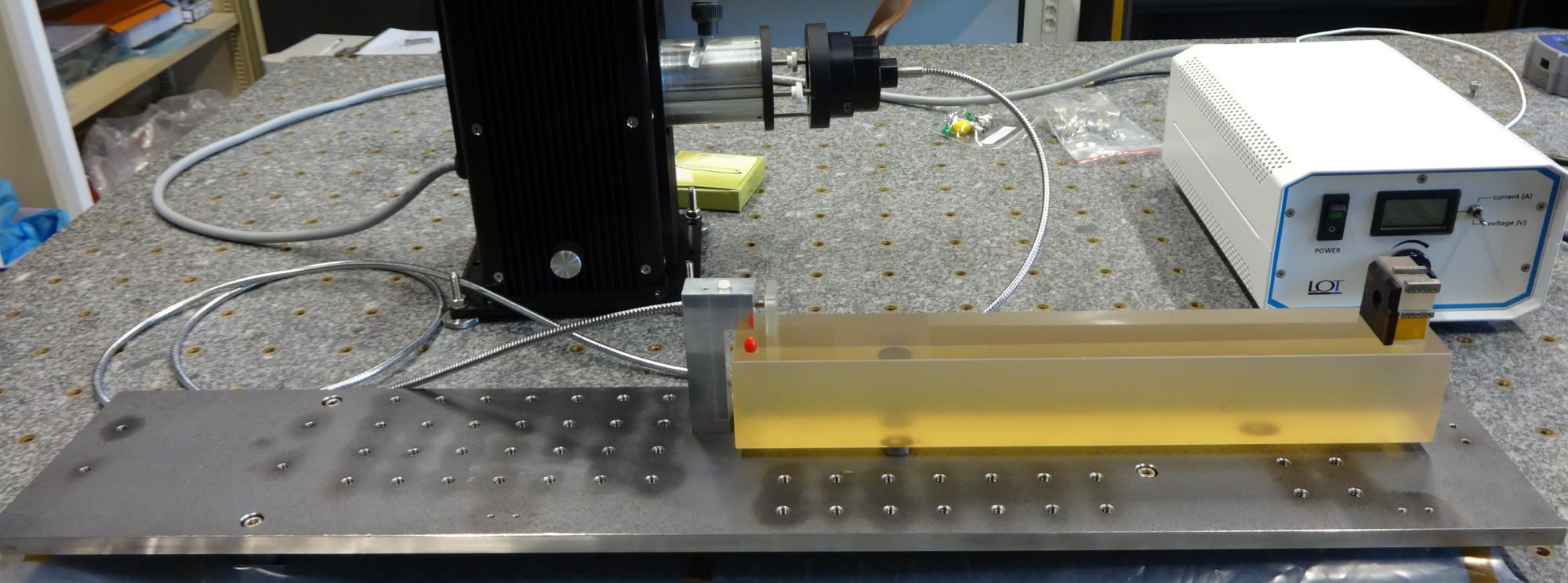}
\caption{\label{fig:invar_and_zerodur_bench}\textbf{The invar and zerodur benches supporting the pseudo stellar sources}. The lower metal bench is made of invar (low dilatation material). The translucent yellowish glass block is the zerodur bench and has an ultra-low dilatation coefficient lower than $1\e{-7} \ml{K^{-1}}$. On the left side of the zerodur bench: the pinhole mask that is back lighted by the liquide core fibre. On the right side: the mirror block with the diaphragm (in black). On can also see metal rails on the top of the mirror block, their function is to hold the metrology fibres in the V-grooves located between them.}
\end{center}
\end{figure}

The pseudo stellar sources system function is to project 5 stars unto the CCD. The 4 outer stars represent reference stars, the central star is the target. This star configuration allows us to perform a precise differential measurement: XY position offset and scale changes can be measured between the stars. The characteristics of this system are summed-up in the Table~\ref{tab:Pseudo stellar sources design}. The light source for the pseudo stellar sources is a black body with a temperature of 3000 K. The goal here is to use a reasonable approximation of the spectrum of a real star, that is why we use white light. The schematic of the system is presented in fig. \ref{fig:scheme_pseudostellar_sources}. Figure \ref{fig:invar_and_zerodur_bench} is a picture of the benches supporting the system.


\begin{table}[t]
\caption{\label{tab:Pseudo stellar sources design}Pseudo stellar sources design.}
\begin{center}
\begin{tabular}{|l|l|l|}
  \hline
  Characteristic & Notation & Value\\
  \hline
 Pupil size $= \frac{\lambda L}{2e}$ & $D$ & $5.0$ mm \\
Magnification & $\gamma$ & $2.0$\\
Sources separation & $s$ & $240$ µm\\
Off-axis angle & - & $2\deg$\\
Bench length & $L$ & $0.60$ m\\
Focal length & $f$ &  $0.20$ m\\
\specialcell{Distance OA\\(pseudo stellar sources to mirror)} & $L_s $ & $0.30$ m\\
  \hline
\end{tabular}
\end{center}
\end{table}

We used a magnification factor of 2 and an off axis angle of 2 degrees. This configuration allows the installation of the pseudo stellar sources and the camera without any beam obstruction with some margins to accommodate the support elements. Additionally, with an aperture as small as 5 mm, a spherical surface is sufficient to obtain optical aberrations that produce a spot diagram smaller than the diffraction pattern whatever its position in the field of view.

\subsection{CCD and electronics}\label{subsec:CCD and electronics}

We chose to use the CCD 39-01 from e2v to take advantage of its high frame rate $\times$ quantum well size, and despite its small matrix size. The characteristics of the CCD (and its electronics) are summarized in Table~\ref{tab:CCD characteristics}.

\begin{table}[t]
\hspace{1cm}
\caption{\label{tab:CCD characteristics}CCD and electronics characteristics.}
\begin{center}
\begin{tabular}{|l|l|}
  \hline
  Characteristic & Value\\
  \hline
Pixel size & $24$ µm\\
Matrix size & $80$x$80$ pixels\\
Read noise & 20 electrons\\
Dark current & 5000 electrons/s @300K\\ 
Effective well size & $200 000$ electrons\\
Read time & 1 ms ($=1$kHz frame rate)\\
Sensitivity range & 0.3 to 0.9 µm\\
  \hline
\end{tabular}
\end{center}
\end{table}

Because the CCD is read at 1kHz, the Poisson noise from the dark current at ambient temperature is only 5 electrons / pixel, which is smaller than the readout noise of 20 electrons. This CCD has two buffer zones (masked pixels) that allow a rapid transfer and high frame rate. The high quantum well size and frame rate allow a very fast integration which is highly desirable given the level of photon noise targeted. The performances expected with this detector are detailed in section \ref{sec:EXPECTED PERFORMANCES AND ERROR BUDGET}.

\section{EXPECTED PERFORMANCES AND ERROR BUDGET}\label{sec:EXPECTED PERFORMANCES AND ERROR BUDGET}

In this section are presented the performance models of the NEAT test bench as well as its preliminary performance budgets.

\subsection{General description}

\begin{figure}[t]
\begin{center}
\includegraphics[width = 150mm]{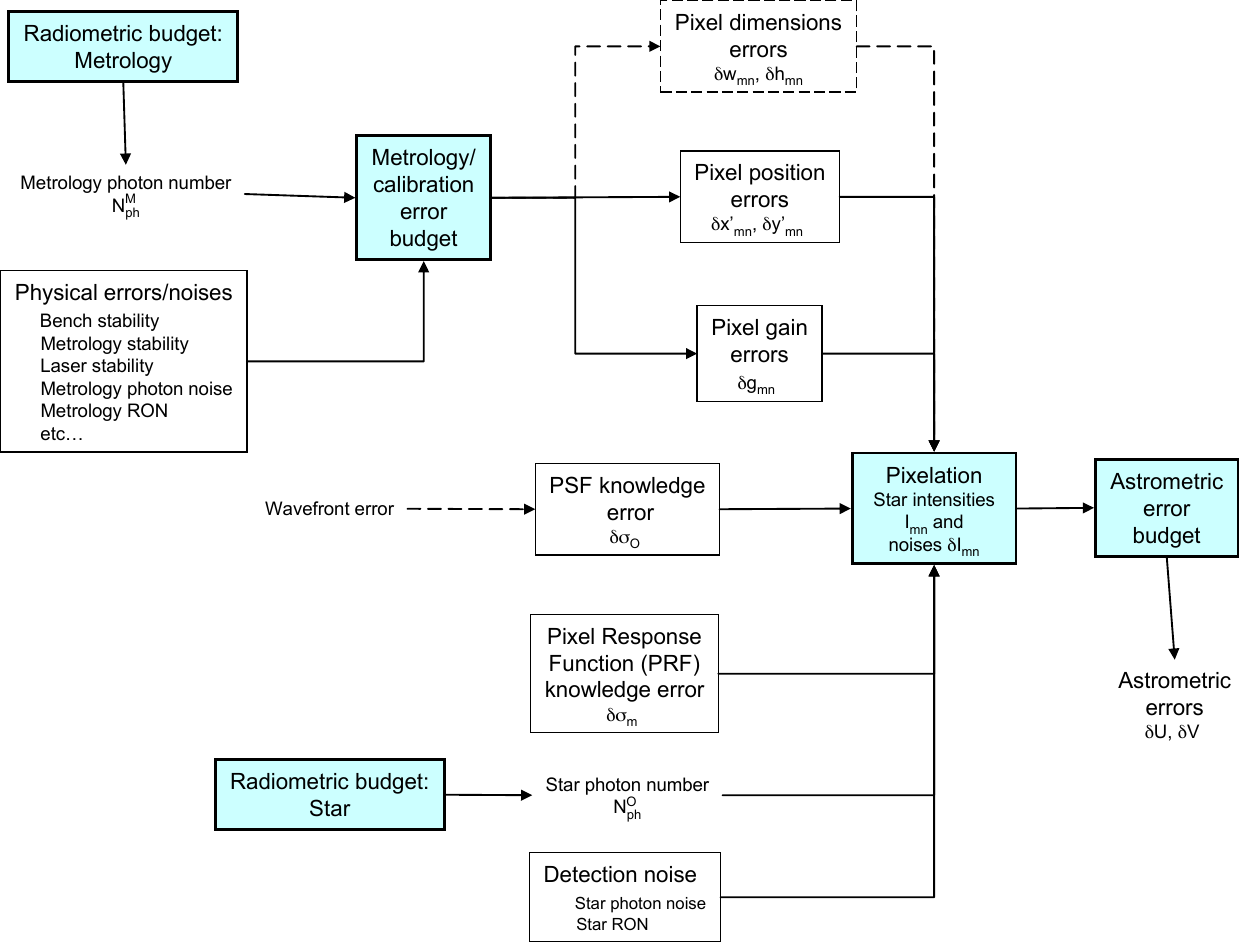}
\caption{\label{fig:errors_budgets}\textbf{General flow-chart of NEAT test bench performance model.}}
\end{center}
\end{figure}

The performance model of the NEAT test bench consists of four separate computation models, as illustrated by the blue boxes in Figure \ref{fig:errors_budgets}. Those models are linked together and are the following:

\begin{enumerate}
\item A radiometric performance model determining the number of photons $N^{\ml{O}}_{\ml{ph}}$ originating from the star simulator and collected by one pixel on the CCD detector.
\item Likewise, a radiometric performance model providing the photons number $N^{\ml{M}}_{\ml{ph}}$ originating from the metrology laser subsystem and collected by one CCD pixel.
\item A ‘‘metrology" (or ‘‘calibration") error model estimating the calibration errors on the pixel widths $w_{mn}$, pixel heights $h_{mn}$, pixel position errors $(x'_{mn}, y'_{mn})$ and pixel gains $g_{mn}$, as a function of various intensity or phase noise terms generated by the metrology system (m and n indicating pixel locations on in the CCD plane). Here m and n are the pixel indices, assuming a M x N detector chip (1 $\leq$ m $\leq$ M  and  1 $\leq$ n $\leq$ N). 
\item A final ‘‘star position" error model, allowing to estimate the astrometric angular errors $\delta U$, $\delta V$ resulting from all the previous parameters, and from the knowledge uncertainties about the telescope Point Spread Function (PSF) and of the Pixel Response Functions (PRF) on-chip.
\end{enumerate}

The photometric models n° 1 and 2 have already been discussed in a previous publication\cite{Crouzier12}, hence it is only dealt here with the calibration and astrometric performance models.

\subsection{Metrology/calibration model}

Assuming an ideal PRF (i.e. uniformly equal to unity for each pixel), we can derive an approximate, first-order expression of the interferograms generated on each pixel (m,n) of the CCD camera by the metrology subsystem, that is:

\begin{equation}\label{eq:Imn stars}
I_{mn} = \frac{N^M_{\ml{ph}} g_{mn}}{\ml{MN}} \left[ 1+V_{mn}\;\ml{sincard}\left(\frac{\pi B_x w_{mn}}{\lambda_m L}\right) \ml{sincard}\left(\frac{\pi B_y h_{mn}}{\lambda_m L}\right) \left[ \cos\left(2\pi\frac{B_x x'_{mn}+B_y y'_{mn}}{\lambda_m L} + \psi(t)\right) \right]\right]
\end{equation}

where $\ml{sincard}(x)$ is the sine cardinal function $x\mapsto\frac{\sin(x)}{x}$. Most of the other scientific notations have already been defined before. In addition,  $V_{mn}$ stands for the averaged visibility of the interferogram on each pixel, $B_X$, $B_Y$ are the baselines between the metrology fibers along the X and Y axes, and $\psi (t)$ is the temporal phase shift introduce between both the metrology system arms. Differentiating this formula, we can introduce various types of error and noises on $N^{\ml{M}}_{\ml{ph}}$ (named "intensity noises") and other parameters such as BX, BY,  lambda m, L and $\psi (t)$ (named "phase noises"), in order to determine the pixels gain and position measurement errors delta gmn, delta x'mn and delat y'mn. The sums of these uncertainties are finally added in RMS sense and presented in Table \ref{tab:errors_calibration_summary_table}. It shows that the most critical error parameters are, by decreasing order of magnitude:

\begin{itemize}
\item Metrology chain shot noise (related to the available power and integration time)
\item Phase control or phase fitting accuracy
\item Stray-light originating from the metrology chain
\end{itemize}

The Table also shows that one of the main goal of the experiment, that is to measure pixel position errors within an accuracy of $5\e{6}$ is not met. However no definitive conclusion can be made at this stage, since we need to estimate their impact on the astrometric performance (this will be the purpose of the next sub-section). Finally, it must be noted that certain error contributions (laser intensity drifts, CCD non-linearity) were set to zero because it is assumed that they can be corrected via data processing.

\begin{table}[h]
\centering
\caption{Error parameters and preliminary results of the calibration budget.}
\includegraphics[width = 160mm]{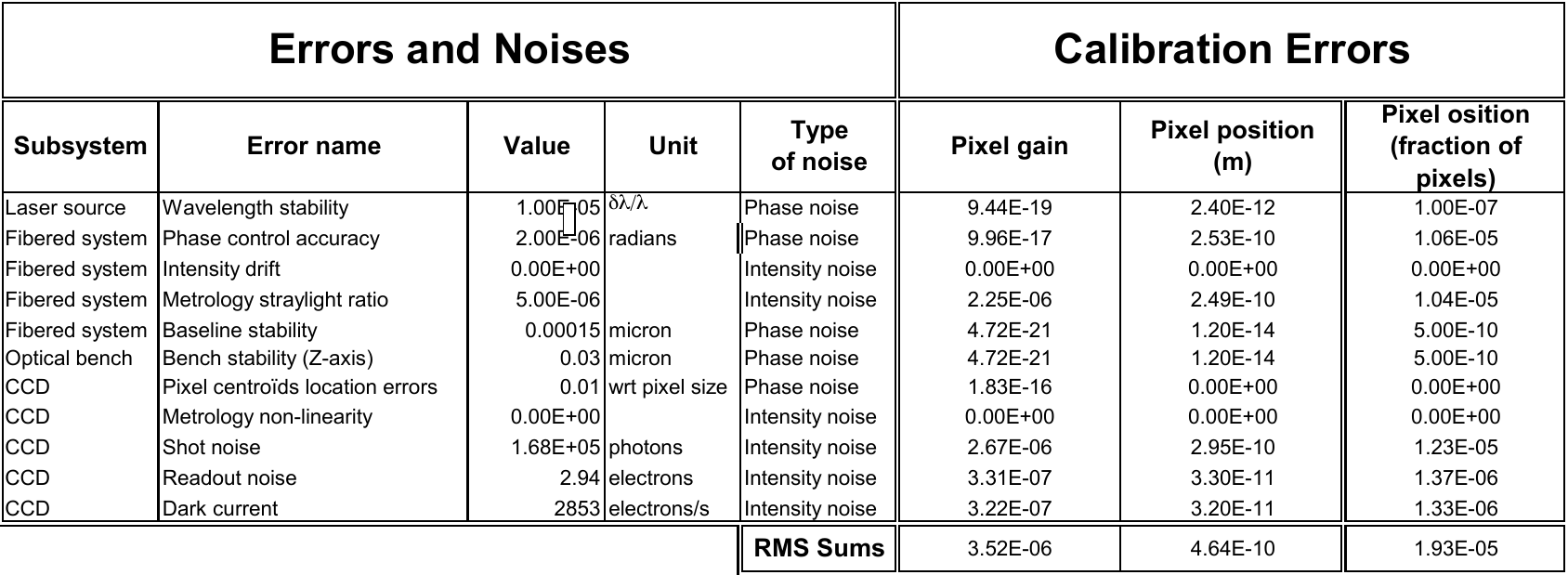}
\label{tab:errors_calibration_summary_table}
\end{table}

\subsection{Astrometric error model}

Using the same types of scientific notations than in the previous sections, we drop out the second spatial dimension directed along the Y-axis for the sake of simplicity – double indices (m,n) being replaced with a single index m – and obtain a general expression of the flux Im integrated along a mono-dimensional pixel along the X-axis: 

\begin{equation}\label{eq:Imn metrology}
I_m = g_m N^{\ml{O}}_{\ml{ph}} \frac{\int_{-w/2}^{w/2} \ml{PSF}(x') \ml{PRF}(x'-x'_m)dx'}{\int_{-\infty}^{+\infty} \ml{PSF}(x')dx'}
\end{equation}

Deriving a simple analytical relationship from Eq. 5 is only feasible at the price of two drastic approximations:

\begin{itemize}
\item It is assumed that PSF($x'$) and PRF($x'$) functions both follow Gaussian distributions whose variances are respectively noted $\sigma_O$ and $\sigma_m$.
\item It is assumed that the term PSF($x'$) can be considered as constant on the pixel width $w$, which allows to remove it out of the integral.
\end{itemize}

Under those hypotheses, Eq. \ref{eq:Imn metrology} can be simplified as:

\begin{equation}\label{eq:simple Imn metrology}
I_m = g_m N^{\ml{O}}_{\ml{ph}} \frac{\sigma_m}{\sigma_O}\exp{\left[\frac{-x'^2_m}{\sigma^2_O}\right]}\ml{Erf}\left[\frac{w}{2\sigma_m}\right]
\end{equation}

where Erf is the error function $\ml{Erf}(x) = \frac{2}{\pi}\int_0^x \ex{-u^2}du$ . In order to estimate the final astrometric error $\delta \overline{x'}$ of the star position resulting from the uncertainties on all the parameters in Eq. 6, we simply use the classical centroïd formula:

\begin{equation}\label{eq:barycenter}
\overline{x'} = \frac{\sum_{m=1}^{M} x'_m I_m}{\sum_{m=1}^{M} I_m}
\end{equation}

Differentiating successively Eq. \ref{eq:simple Imn metrology} then \ref{eq:barycenter} allows to list the major error sources that are summarized on the left side of Table \ref{Error parameters for the star position budget}. Here the pixels gains and positions errors are directly related to those estimated from the calibration budget presented in the last sub-section. The other terms essentially are uncertainties about pixels width, knowledge of PSF half-width $\sigma_O$, and knowledge of PRF half-width $\sigma_m$.

\begin{table}[t]
\centering
\caption{\label{Error parameters for the star position budget}\textbf{Error parameters for the star position budget.} (a): Pixel/PSF/PRF knowledge errors. (b) Centroids errors.}
\subfigure[]{\label{tab:knowledge_errors_centroids_parameters_table}
\includegraphics[width=60mm]{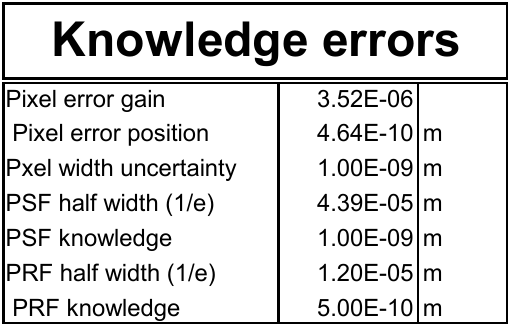}}
\hspace{5pt}
\subfigure[]{\label{tab:cerrors_centroids_summary_table}
\includegraphics[width=90mm]{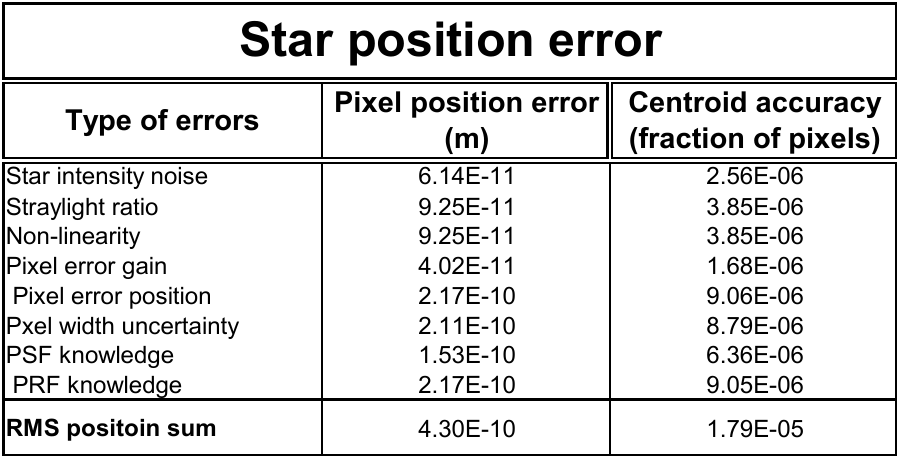}}
\end{table}

The Table \ref{tab:cerrors_centroids_summary_table} is presented a preliminary error budget for the star position measurement accuracy. It shows that the main contributors are by decreasing order of magnitude the pixel position measurement errors, the uncertainty on their actual width, and on their PRF knowledge. We finally see that the actual astrometric performance is around $2\e{-5}$ pixels, which is four times greater than the original requirement of $5\e{-6}$. However, it must be emphasized that some error items featuring in Table  \ref{tab:knowledge_errors_centroids_parameters_table} (e.g. pixel width, PSF and PRF knowledge uncertainties) should certainly be minimized by means of additional calibration processes or data processing refinements. This will be the scope of future works.

\section{CURRENT STATUS OF THE TESTBED AND LAST RESULTS}\label{sec:CURRENT STATUS OF THE TESTBED AND LAST RESULTS}

\subsection{Current status}

To date, all components have been procured. The assembling of the bench is under-way and will soon be completed. The first images have been produced, in air. The first light in air was obtained in July 2013, for both the metrology fringes and the pseudo stars, constituting two sets of data cubes. The next step will be to obtain the same set of images, in vacuum.

\subsection{Metrology results}

\begin{figure}[!t]
\centering
\subfigure[]{\label{fig:vertical_fringes}
\includegraphics[width=70mm]{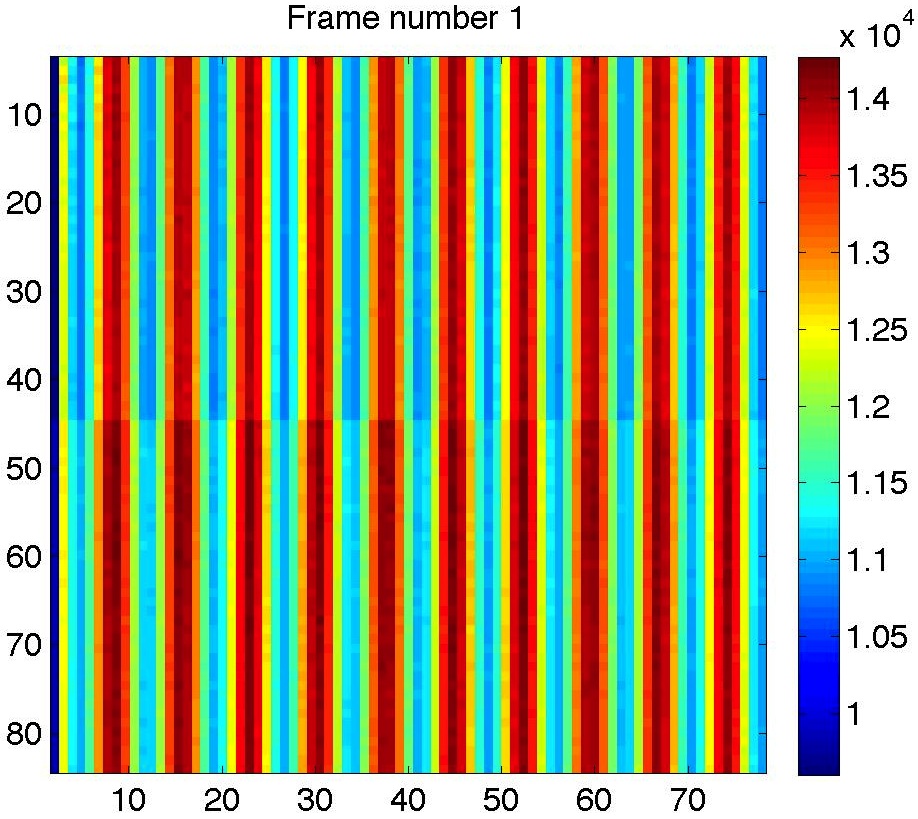}}
\hspace{5pt}
\subfigure[]{\label{fig:horizontal_fringes}
\includegraphics[width=70mm]{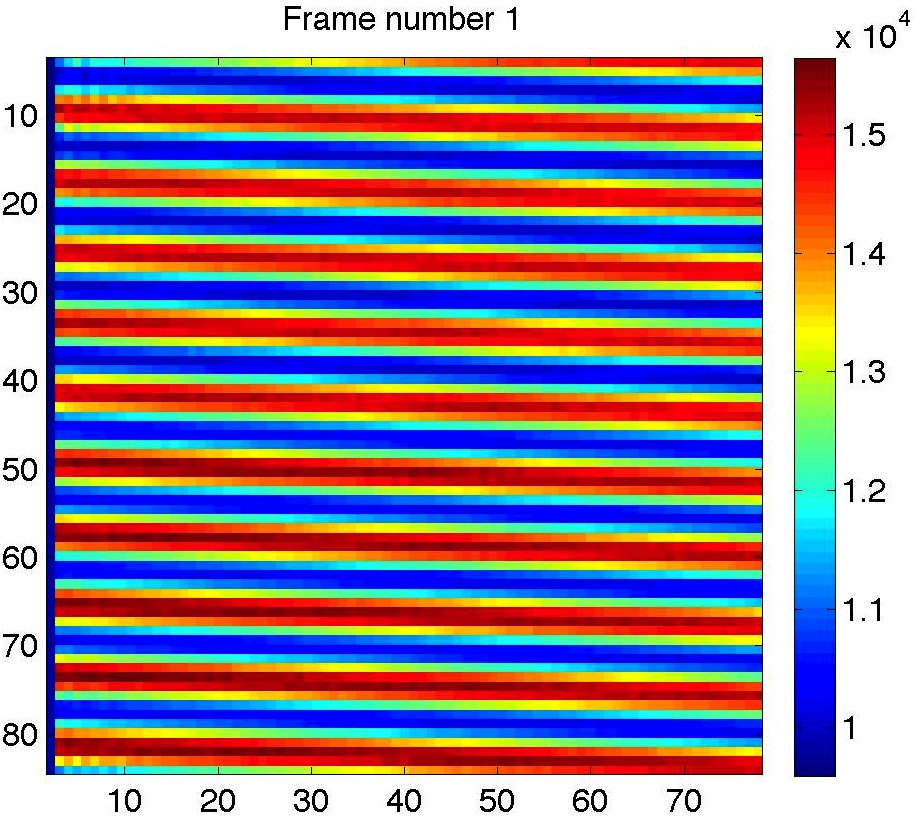}}
\caption{\label{fig:metrology_fringes}\textbf{Vertical (a) and horizontal (b) metrology fringes.} 12 minutes of data acquired at 100Hz resulted in a data cube with 70000 usable frames. The visibility of the fringes V=$\frac{A_{\ml{max}}-A_{\ml{min}}}{A_{\ml{max}}+A_{\ml{min}}}$ after dark subtraction is 0.64 and this value is stable with time. This result is important because it validates the design of the metrology: it is important to have a stable and fairly high visiblity. The useful photons for the measure of the positions are the ones that interfere, the constant background level only adds noise.}
\end{figure}

The first set of data cubes, the metrology runs, is shown by fig. \ref{fig:vertical_fringes} and \ref{fig:horizontal_fringes}. The fringes are dynamic because the phase is modulated in each lane of the baselines. The modulation applied on each lane was a triangle signal of amplitude $10$V (out of phase between the lanes). This resulted into a back and forth, constant velocity sweep motion with an amplitude of about 3 times the inter-fringe distance on the CCD.
\\

\begin{figure}[!h]
\centering
\subfigure[]{\label{fig:deltay_all}
\includegraphics[width=95mm]{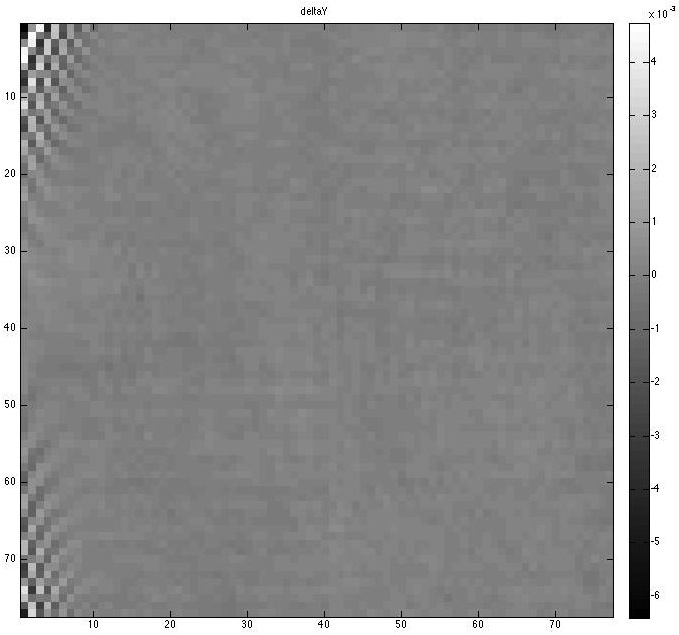}}
\subfigure[]{\label{fig:deltay_right}
\includegraphics[width=70mm]{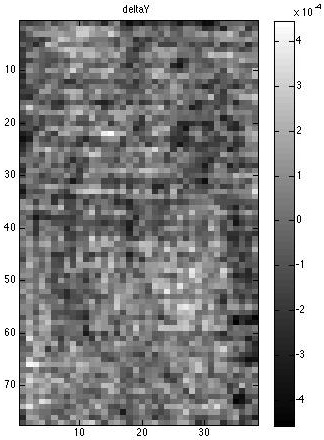}}
\caption{\label{fig:deltay}\textbf{Measured pixel position offsets (in the horizontal direction).} The figure on the left shows the entire CCD. On the left corners of the CCD we see large biases caused by the baffle: its position could not be properly adjusted because the translation stage is being repaired and was replaced by a dummy. The figure on the right shows the right part of the CCD only, where the pixel positions are not affected. The unit of the color scale is hundred of micropixels: the pixels offsets scale between -400 and +400 micropixels. The map of the position offsets in the vertical direction as very similar properties.}
\end{figure}

\begin{figure}[!h]
\begin{center}
\includegraphics[width = 160mm]{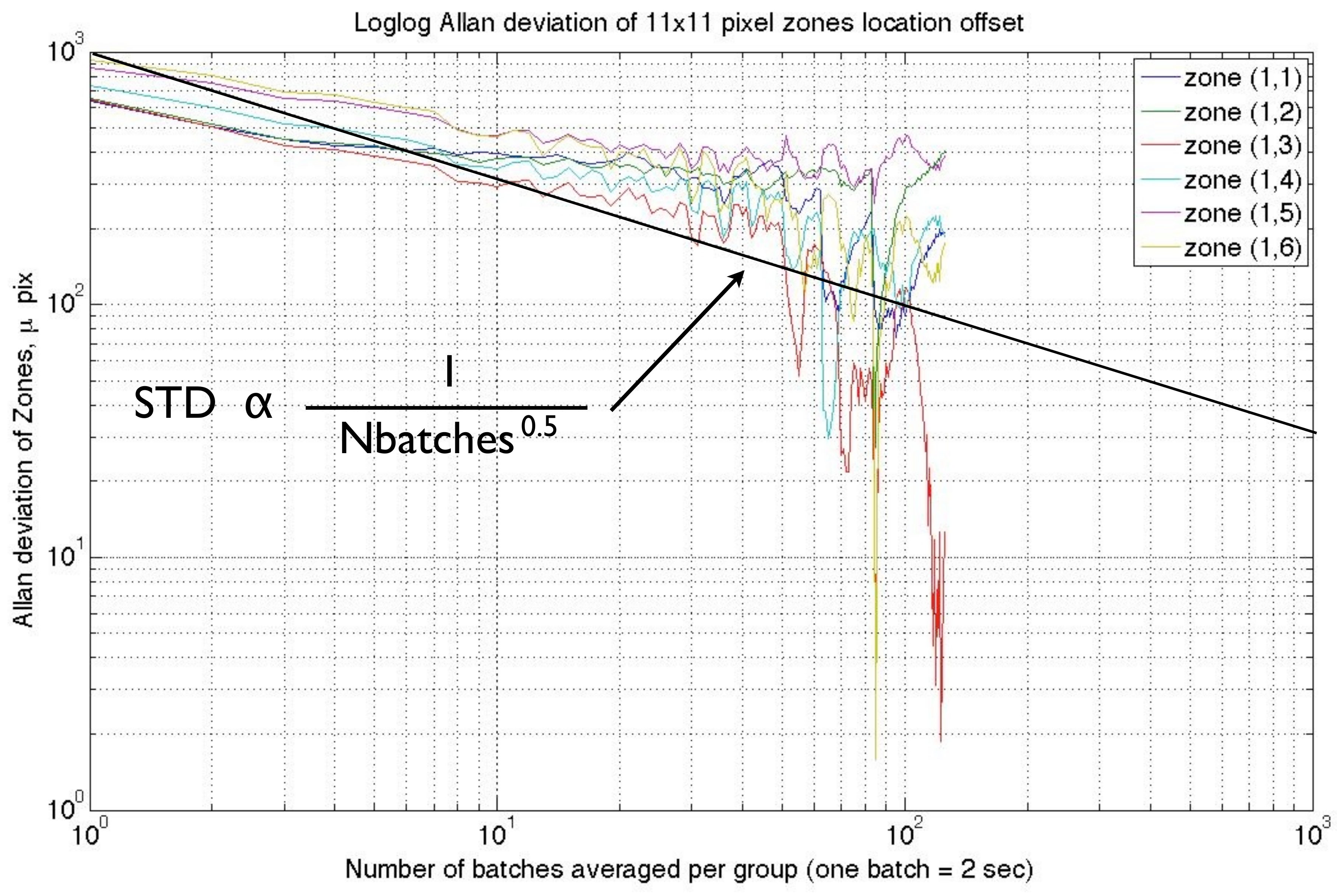}
\caption{\label{fig:semilog_allan_deviation_zones}\textbf{Allan deviations of the position of groups of pixels.}}
\end{center}
\end{figure}

This data was processed in order to determine the pixels location offsets, the results are presented by fig. \ref{fig:deltay} and \ref{fig:semilog_allan_deviation_zones}. The pixel positions are obtained by looking at the phase difference between the global system of fringes and the signal given by individual pixels. The detail of the method used to find the pixel positions is not presented here.
\\

Figure \ref{fig:semilog_allan_deviation_zones} show the \emph{Allan deviation} of zones of pixels (of 11 by 11 pixels). The frames are grouped into batches because we need several frames to measure the pixels positions. We have for each batch a measure of each pixel position offset. From this we can calculate the Allan deviation of the position offset of each pixel: we form group of batches. Within each group we average the positions given by the batches and we calculate the deviations of the offsets between all the groups. We obtain the final plot by varying the number of batches in each group. The precision increases as we use groups with more batches (and more frames), thus reducing the photon noise. To reduce the photon noise even further, we also average spatially over zones of pixels. The end of the curve approaches the precision of the measurement. The black curve shows the theoretical slope for pure white noise: in this case the precision should increase with the power -0.5 of the number of frames.

\subsection{Pseudo stars results}

\begin{figure}[!h]
\centering
\subfigure[]{\label{fig:stars_pos1}
\includegraphics[width = 60mm]{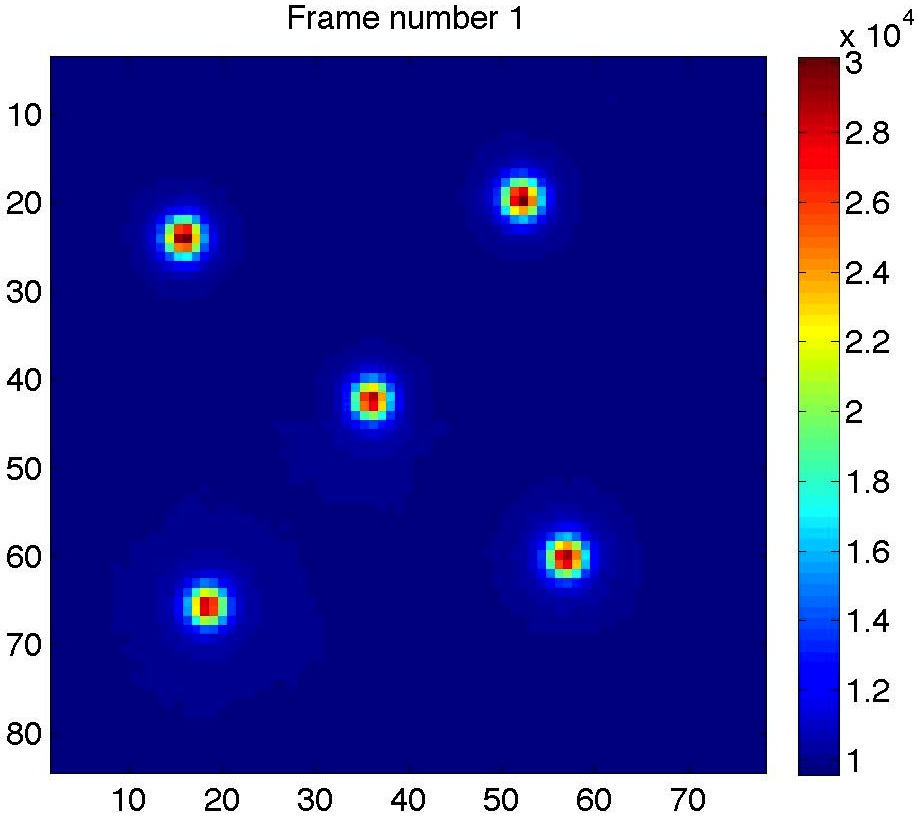}}
\hspace{5pt}
\subfigure[]{\label{fig:stars_pos2}
\includegraphics[width = 60mm]{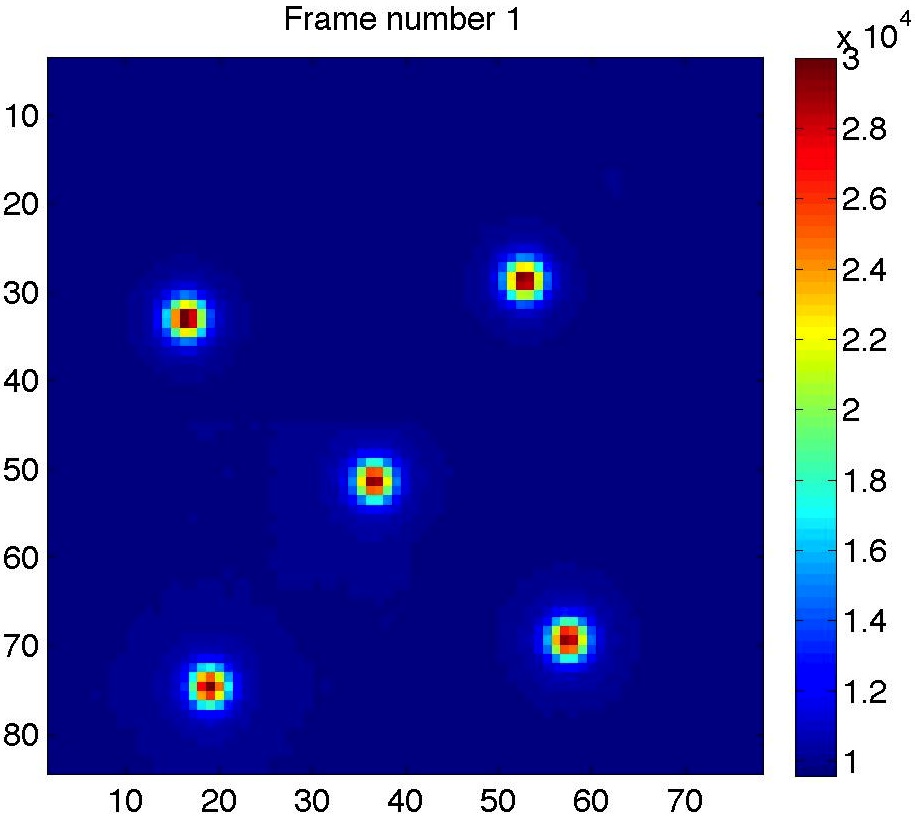}}
\subfigure[]{\label{fig:stars_pos3}
\includegraphics[width = 60mm]{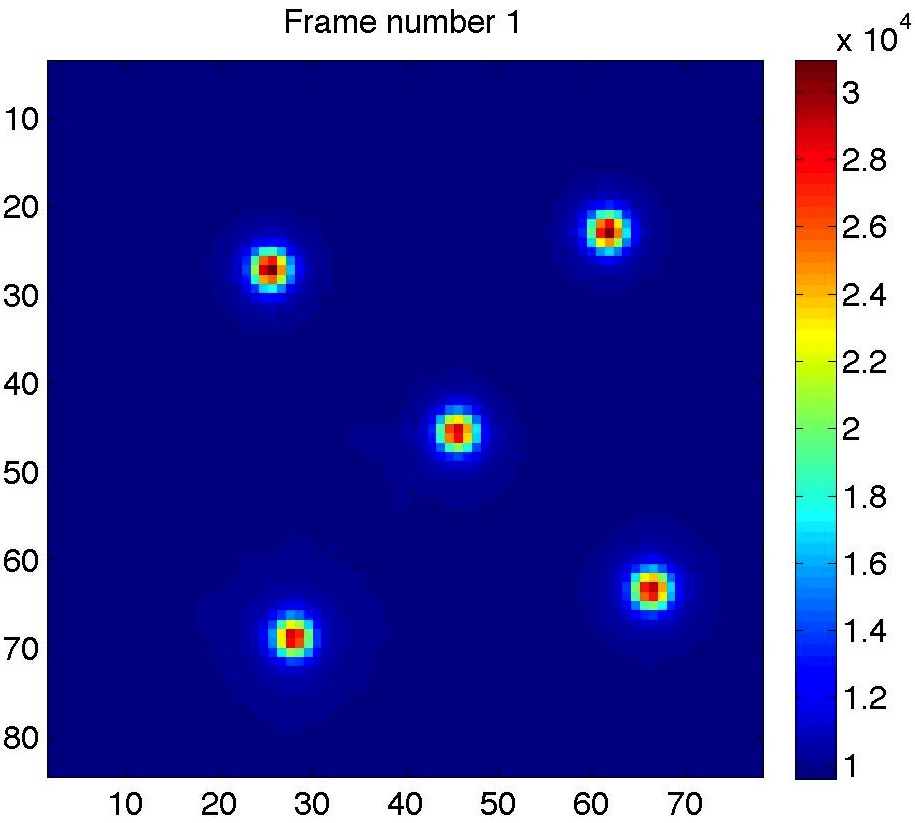}}
\hspace{5pt}
\subfigure[]{\label{fig:stars_pos4}
\includegraphics[width = 60mm]{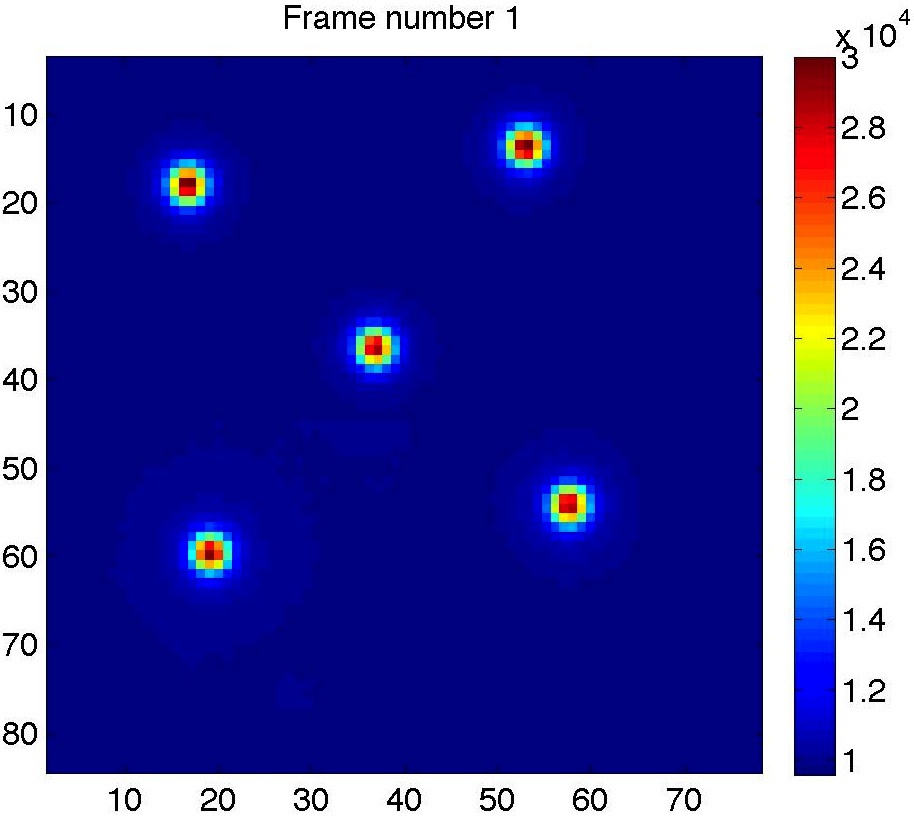}}
\caption{\label{fig:stars_pos}\textbf{Images of the pseudo stars at 4 different positions on the CCD.}}
\end{figure}

The second set of data cubes, the pseudo stars runs, are shown by fig. \ref{fig:stars_pos1}, \ref{fig:stars_pos2}, \ref{fig:stars_pos3}, and \ref{fig:stars_pos4}. With the pseudo stars data, we have performed two types of analysis: either by using only one of the four runs (so the position of the centroids is fixed), or by using all four.

\begin{figure}[!h]
\centering
\subfigure[]{\label{fig:centroid_positions}
\includegraphics[width=70mm]{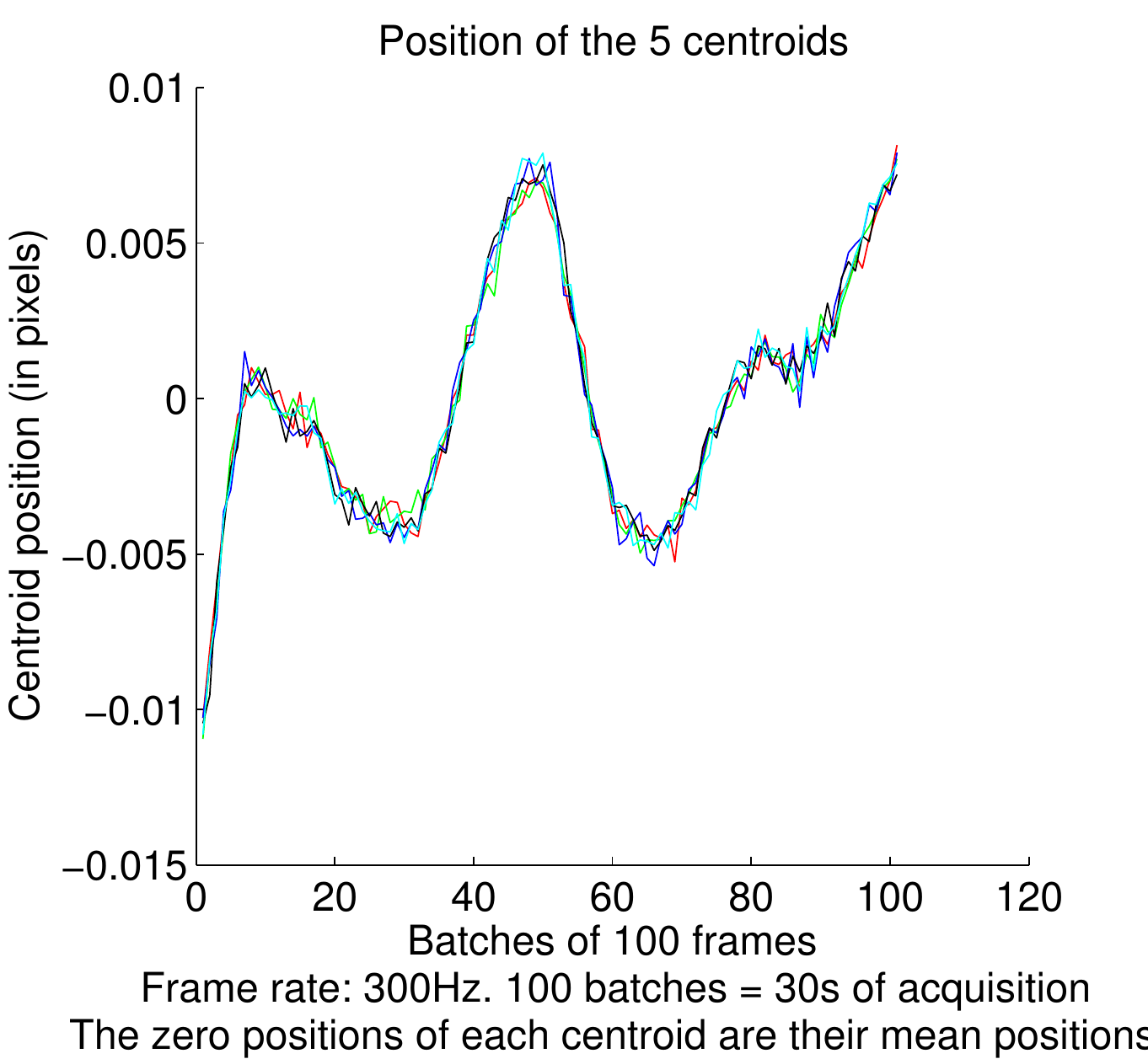}}
\hspace{5pt}
\subfigure[]{\label{fig:centroid_distance_std}
\includegraphics[width=70mm]{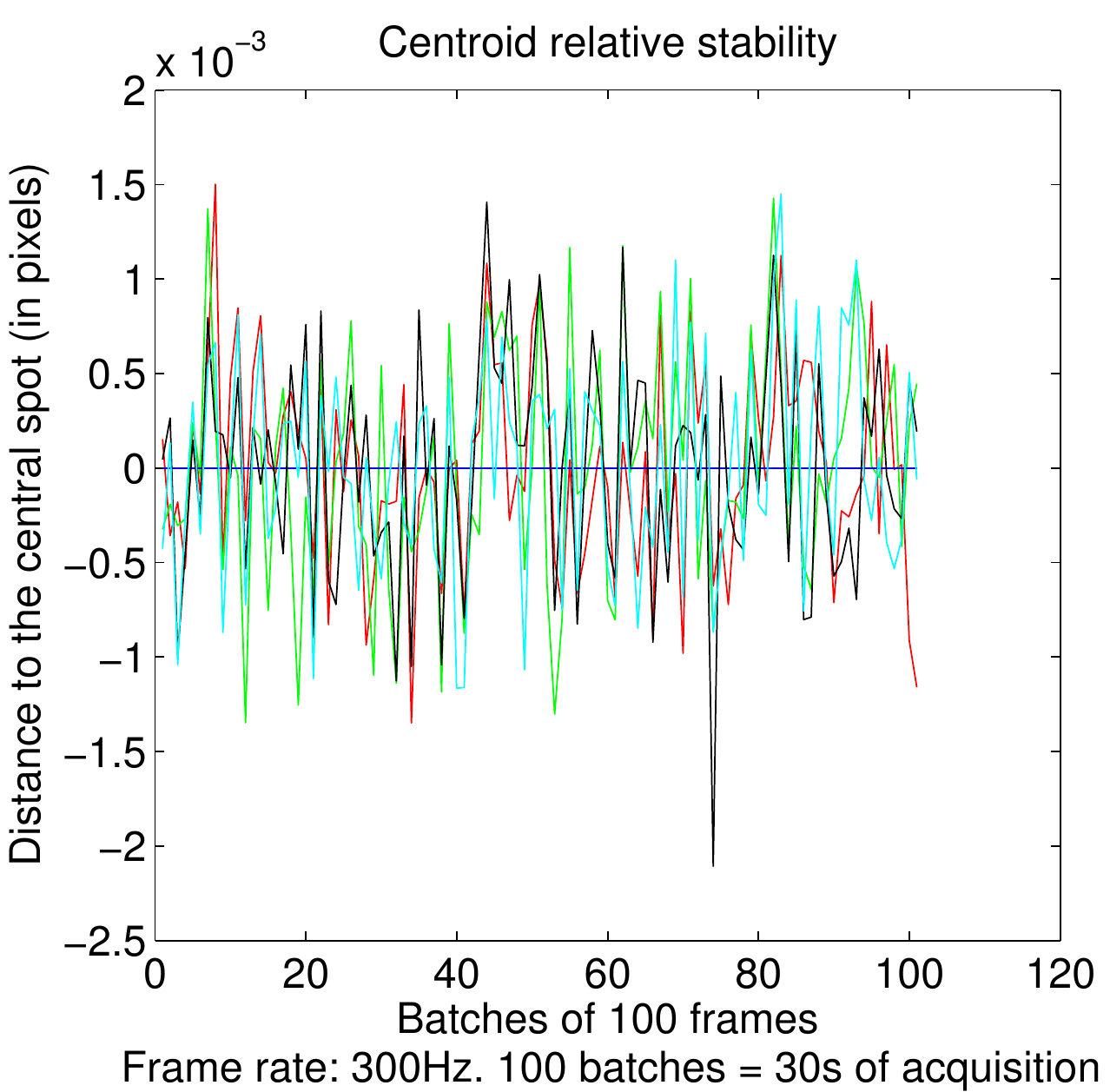}}
\caption{\label{fig:pseudo_stars_run}\textbf{Absolute and relative centroid positions measured for a fixed CCD position.} The centroiding method used here is a simple Gaussian fit. The pixels positions from the metrology were not taken into account. The standard deviation for each centroid is about 500 micro pixels.}
\end{figure}

For the first analysis, we measure the centroids locations for one of the four positions only. This measure is sensible to some environmental factors such as mechanical stability, air turbulence etc... but if the centroid positions are stable enough, the pixel errors are constant and therefore do not affect the measures significantly. We can measure the distances between the centroids versus the time with a very high stability (standard deviation of 500 micro pixels). The location of the centroids versus the time (in the CCD frame) is given by fig. \ref{fig:centroid_positions}. The distances variations between the centroids are given by fig. \ref{fig:centroid_distance_std}.

For the second analysis, we measure the distances between the centroids locations, for each one of the four positions (i.e. data cube). We can then calculate the deviation of the distances between the outer centroids and the central one. The results are standard deviations (one for each outer centroid) calculated over only 4 values. The final result has huge error bars because four positions is a very small sample. This is because we had to manually moved the CCD between each run, the translation stage being unavailable. The values obtained were on the order of dozens of mili pixels. This means that when the centroids are moving, systematics errors most likely due to the pixelation of the centroids and the simple fitting procedure become dominant.

\section{CONCLUSION}

We are in the process of assembling a testbed that will demonstrate the feasibility of measuring centroids to a precision of $5\e{-6}$ pixel. This will strengthen the case for NEAT as it will show that astrometry down to sub-microarcsec precision is a valid technique for searching Earth-like exoplanets in the habitable zone of nearby stars.

To this day, all the components have been procured, the assembling of the bench is to be completed soon and the first images have been obtained in the air. So far, the quality of the first images validates the design. A precision of 100 micro pixels was obtained on the measurement of the mean location offset of a group of pixels.

The results we have presented here are only a first attempt. They are numerous actions that will be taken in order to improve the precision: we are going to put the experiment in vacuum, to stabilise and cool the CCD temperature from ambient room temperature to slightly below 0$\dg$C, to compare the precision of various ways of fitting the centroids and to use the information from the metrology to correct for pixel location offsets and widths. 

An error budget of the experiment has also been presented. This tool will be very useful to identify and mitigate the dominant sources of noise present in the data, in order to reach the targeted precision.


\acknowledgments     

We would like to thank the engineering team at IPAG for their support. We acknowledge the labex OSUG@2020 and CNES for financing the experiment and CNES and Thales Alenia Space for funding the PhD of A.\ Crouzier. At last, we are grateful to Bijan Nemati, Chengxing Zhai and Inseob Hahn for hosting Antoine Crouzier into their team and taking the time to answer many of our questions. 


\bibliography{spie_article_antoine_crouzier_2013}   

\begin{thebibliography}{1}

\bibitem{Malbet11}
{Malbet}, F., {L{\'e}ger}, A., {Shao}, M., {Goullioud}, R., {Lagage}, P.-O.,
  {Brown}, A.~G.~A., {Cara}, C., {Durand}, G., {Eiroa}, C., {Feautrier}, P.,
  {Jakobsson}, B., {Hinglais}, E., {Kaltenegger}, L., {Labadie}, L.,
  {Lagrange}, A.-M., {Laskar}, J., {Liseau}, R., {Lunine}, J., {Maldonado}, J.,
  {Mercier}, M., {Mordasini}, C., {Queloz}, D., {Quirrenbach}, A., {Sozzetti},
  A., {Traub}, W., {Absil}, O., {Alibert}, Y., {Andrei}, A.~H., {Arenou}, F.,
  {Beichman}, C., {Chelli}, A., {Cockell}, C.~S., {Duvert}, G., {Forveille},
  T., {Garcia}, P.~J.~V., {Hobbs}, D., {Krone-Martins}, A., {Lammer}, H.,
  {Meunier}, N., {Minardi}, S., {Moitinho de Almeida}, A., {Rambaux}, N.,
  {Raymond}, S., {R{\"o}ttgering}, H.~J.~A., {Sahlmann}, J., {Schuller}, P.~A.,
  {S{\'e}gransan}, D., {Selsis}, F., {Surdej}, J., {Villaver}, E., {White},
  G.~J., and {Zinnecker}, H., ``High precision astrometry mission for the
  detection and characterization of nearby habitable planetary systems with the
  nearby earth astrometric telescope (neat),'' {\em Experimental Astronomy} ,
  109 (Sep 2011).

\bibitem{Malbet12}
Malbet, F., Goullioud, R., Lagage, P., Léger, A., Shao, M., Crouzier, A., and
  consortium NEAT, ``Neat: a space born astrometric mission for the detection
  and characterization of nearby habitable planetary systems,'' {\em Proc. of
  SPIE}~{\bf 8442} (2012).

\bibitem{Malbet13}
Malbet, F., Crouzier, A., Léger, A., and al., ``Neat: an astrometric mission
  to detect nearby planetary systems down to the earth mass,'' {\em Proc. of
  SPIE}~{\bf 8864} (2013).

\bibitem{sim_double_blind_test09}
Traub, W., Ford, E., Laughlin, G., Levison, H., Lin, D., Raymond, S., Makarov,
  V., Casertano, S., Fischer, D., Kasdin, J., Muterspaugh, M., Shao, M.,
  Beichman, C., Boss, A., Gould, A., and Marr, J., ``Overview of the sim-rv
  double-blind simulation to detect earths in multi-planet systems,'' in [{\em
  American Astronomical Society Meeting Abstracts
  \#213}{\nolinebreak\hspace{0.1em}]},  {\em Bulletin of the American
  Astronomical Society} {\bf 41},  \#300.01 (Jan 2009).

\bibitem{neat_number_of_measurements11}
Malbet, F. and Léger, A., ``Neat document: number of visits needed per
  target,'' Private communication (2011).

\bibitem{neat_error_budget11}
Goullioud, R., ``Neat error budget,'' Private communication (2011).

\bibitem{Nemati11}
Nemati, B., Shao, M., Zhai, C., Erlig, H., Wang, X., and Goullioud, R.,
  ``Micro-pixel image position sensing testbed,'' {\em Proc. of SPIE}~{\bf
  8151} (2011).

\bibitem{Zhai11}
Zhai, C., Shao, M., Goullioud, R., and Nemati, B., ``Micro-pixel accuracy
  centroid displacement estimation and detector calibration,'' {\em Royal
  Society of London Proceedings Series A}~{\bf 467},  3550--3569 (2011).

\bibitem{Crouzier12}
Crouzier, A., Malbet, F., Preis, O., Henault, F., Kern, P., Martin, G.,
  Feautrier, P., Cara, C., Lagage, P.~O., Léger, A., LeDuigou, J.~M., Shao.,
  M., and Goullioud, R., ``Neat: a space born astrometric mission for the
  detection and characterization of nearby habitable planetary systems,'' {\em
  Proc. of SPIE}~{\bf 8445} (2012).

\end{thebibliography}
\bibliographystyle{spiebib}   

\end{document}